\documentclass[usenatbib]{mn2e}
\usepackage{graphicx}
\usepackage{amsmath}

\def\apj{\rm ApJ}
\def\apjl{\rm ApJL}
\def\apjs{\rm ApJS}
\def\aj{\rm AJ}
\def\mnras{\rm MNRAS}

\def\pasa{\rm PASA}
\def\aap{\rm AAP}

\def\physrep{\rm Physics Reports}

\def\gax{\mathrel{\raise.3ex\hbox{$>$}\mkern-14mu\lower0.6ex\hbox{$\sim$}}}
\def\lax{\mathrel{\raise.3ex\hbox{$<$}\mkern-14mu\lower0.6ex\hbox{$\sim$}}}
\def\gtorder{\mathrel{\raise.3ex\hbox{$>$}\mkern-14mu
             \lower0.6ex\hbox{$\sim$}}}
\def\ltorder{\mathrel{\raise.3ex\hbox{$<$}\mkern-14mu
             \lower0.6ex\hbox{$\sim$}}}

\begin{document}

\title [Unbound Companions to Pulsar J1124$-$5916]
   {A Search for Unbound Stellar Companions to Pulsar J1124$-$5916}

\author[C.~S. Kochanek]{ 
    C.~S. Kochanek$^{1,2}$
    \\
  $^{1}$ Department of Astronomy, 
       The Ohio State University, 140 West 18th Avenue, Columbus OH 43210 \\
  $^{2}$ Center for Cosmology and AstroParticle Physics, The Ohio State University,
    191 W. Woodruff Avenue, Columbus OH 43210 \\
   }

\maketitle

\begin{abstract}
We searched for and found no higher mass ($\gtorder 3M_\odot$) unbound binary
stellar companions to the progenitor of pulsar J1124$-$5916.  There are
lower mass candidates, but they all have high probabilities of being false
positives.  There are no candidates for it now being a fully unbound
triple system.  Even if one of the lower mass candidates is an unbound
companion, it seems unlikely that it could have contributed to stripping
the progenitor prior to the supernova.  The stars are too low mass to be
significant mass gainers, and they are too slowly moving to be the 
survivors of a compact, post-common envelope binary.  The addition of
one more system slightly improves the statistical constraints on the 
binary and triple status of supernova progenitors just before and
after death.
\end{abstract}


\section{Introduction}

Massive stars are overwhelmingly born in binaries and higher order systems 
(e.g., \citealt{Kobulnicky2007}, \citealt{Sana2012}, \citealt{Moe2017}, \citealt{Villasenor2021})
and this has profound effects on their evolution and ultimate fates.  
Predictions about their deaths in supernovae or implosion, the numbers of interacting
massive compact object binaries, or gravitational wave mergers all require
understanding their multiplicity and the evolution of binary systems, which
can be explored in binary population synthesis models (e.g., 
\citealt{Belczynski2002}, \citealt{Kalogera2007}, \citealt{Belczynski2008},
\citealt{Dominik2013}, \citealt{Antonini2017}, \citealt{Eldridge2017}, 
\citealt{Fragione2019}).  Many of the uncertainties in these models 
for the fates of massive stars could be reduced given direct constraints
on their multiplicity just before and after their deaths in supernovae
rather than only at birth.

The multiplicity of supernovae leading to neutron stars can be
well constrained using the roughly 300 Galactic supernova remnants (e.g.,
\citealt{Green2019}) up to the limitations imposed by extinction. The
original approach (\citealt{vandenBergh1980}, \citealt{Guseinov2005})
was simply to look for O and B stars associated with supernova remnants.
Surviving, bound binaries can be constrained from the optical properties
of the associated compact objects (\citealt{Kochanek2018}).  \cite{Kochanek2019}
examined 23 SNRs with known compact objects, finding three interacting
binaries (the black hole Roche accretion system SS~433 (e.g., \citealt{Hillwig2008})
and the wind accretion systems HESS~J0632$+$057 (\citealt{Hinton2009})
and 1FGL~J1018.6$-$5856 (\citealt{Corbet2011})) and no non-interacting
binaries.  Unbound binaries can be identified by searching for stars 
with proper motions that either intersect the center of the SNR or the
motion of the neutron star at a reasonable look back time (\citealt{Dincel2015},
\citealt{Boubert2017}, \citealt{Fraser2019}, \citealt{Kochanek2021}, \citealt{Lux2021}).
In particular, \cite{Kochanek2021} surveyed 10 SNRs containing neutron stars
with proper motion measurements to find only one unbound binary, the star
HD~37424 in G180.0$-$01.7 that had been previously identified by 
\cite{Dincel2015}. \cite{Kochanek2021} then combines all these available
statistics to provide the first direct measurements of the binary properties
of supernovae producing neutron stars: 72\% are not binaries at death,
14\% are bound binaries and 13\% are unbound binaries after the explosion.
Roughly speaking, these limits are for stars $\gtorder 3M_\odot$, although
individual mass limits can be significantly smaller.  The statistical uncertainties 
are still large, but they are also only based on $\sim 10\%$ of the Galactic SNRs.

Recently, \cite{Long2022} measured a proper motion for the pulsar J1124$-$5916
(\citealt{Hughes2001}, \citealt{Camilo2002}) in the SNR G292.0$+$1.8.  The
separation of the pulsar from the expansion center of the SNR (\citealt{Winkler2009})
already implied a high velocity, and \cite{Long2022} find a velocity of
$612\pm 152$~km/s in the Gaia reference frame.  Relative to the local standard
of rest of the stars, the velocity is somewhat higher (744~km/s, see below).
G292.0$+$1.8 is an oxygen rich SNR where the recent X-ray abundance analysis 
of \cite{Bhalerao2019} suggests a progenitor mass of $13$ to $30M_\odot$,
lower than earlier estimates (\citealt{Gonzalez2003}, \citealt{Yang2014}, 
\citealt{Kamitsukasa2014}).  \cite{Temim2022} argue for an even lower progenitor
mass of $12$-$16M_\odot$.  The explosion has swept up a large amount of 
mass ($\sim 14M_\odot$, \citealt{Chevalier2005}, \citealt{Bhalerao2019})
some of which lies in an ``equatorial ring'' (\citealt{Park2007}, \citealt{Ghavamian2005})
which is interpreted as material lost by the progenitor.  The lower progenitor masses 
combined with the large mass of circumstellar material implies that the 
progenitor must have been stripped by either winds or binary interactions
prior to the explosion.  The equatorial ring is also suggestive of the presence
of a binary.   \cite{Temim2022} model the SNR and are able to reproduce its properties
under the assumption of a stripped progenitor exploding into a large mass of previously
ejected gas.

Here we combine the \cite{Long2022} proper motion measurement with 
Gaia EDR3 (\citealt{Gaia2016}, \citealt{Gaia2021})
to search for disrupted binary or triple stellar companions to
PSR~J1124$-$5916.  We summarize the data and the method developed
to search for companions by \cite{Kochanek2021} in \S2.  In \S3
we present the results, and we discuss them in \S4. 

\begin{table*}
  \centering
  \caption{CMD Selection Limits}
  \begin{tabular}{lcccccccc}
  \hline
  \multicolumn{1}{c}{Mass} &
  \multicolumn{1}{c}{Color} &
  \multicolumn{1}{c}{Mag} &
  \multicolumn{1}{c}{Color)} &
  \multicolumn{1}{c}{Mag} &
  \multicolumn{1}{c}{Color} &
  \multicolumn{1}{c}{Mag} &
  \multicolumn{1}{c}{Color} &
  \multicolumn{1}{c}{Mag} \\
  \hline
  \multicolumn{1}{c}{$>3M_\odot$} &
  \multicolumn{1}{c}{$<0.7$} &
  \multicolumn{1}{c}{$\hphantom{-}1.0$} &
  \multicolumn{1}{c}{$0.7$-$1.1$} &
  \multicolumn{1}{c}{$\hphantom{-}2.0$} &
  \multicolumn{1}{c}{$1.1$-$2.0$} &
  \multicolumn{1}{c}{$-1-3(B_P-R_P-1)$} &
  \multicolumn{1}{c}{$>2$} &
  \multicolumn{1}{c}{$-4$} \\
  \multicolumn{1}{c}{$>4M_\odot$} &
  \multicolumn{1}{c}{$<0.0$} &
  \multicolumn{1}{c}{$\hphantom{-}0.5$} &
  \multicolumn{1}{c}{$0.0$-$1.3$} &
  \multicolumn{1}{c}{$-1.9$} &
  \multicolumn{1}{c}{$1.3$-$2.0$} &
  \multicolumn{1}{c}{$-1-3(B_P-R_P-1)$} &
  \multicolumn{1}{c}{$>2$} &
  \multicolumn{1}{c}{$-4$} \\
  \multicolumn{1}{c}{$>5M_\odot$} &
  \multicolumn{1}{c}{$<0.0$} &
  \multicolumn{1}{c}{$-0.4$} &
  \multicolumn{1}{c}{$0.0$-$1.3$} &
  \multicolumn{1}{c}{$-1.9$} &
  \multicolumn{1}{c}{$1.3$-$2.0$} &
  \multicolumn{1}{c}{$-1-3(B_P-R_P-1)$} &
  \multicolumn{1}{c}{$>2$} &
  \multicolumn{1}{c}{$-4$} \\
  \hline
\multicolumn{9}{l} {
  Mag gives the upper limit on $M_G$ for each $B_P-R_p$ color range. }
  \end{tabular}
  \label{tab:maglim}
\end{table*}

\begin{table*}
  \centering
  \caption{Searching For Disrupted Binaries}
  \begin{tabular}{crrrrrrrrrr}
  \hline
  \multicolumn{1}{c}{$M_{lim}$} &
  \multicolumn{1}{c}{$N_*$} &
  \multicolumn{3}{c}{$90\%$} &
  \multicolumn{3}{c}{$95\%$} &
  \multicolumn{3}{c}{$99\%$} \\
  \multicolumn{1}{c}{($M_\odot$)}
   &
  &\multicolumn{1}{c}{$N_{obs}$}
  &\multicolumn{1}{c}{$N_{ran}$}
  &\multicolumn{1}{c}{Prob.}
  &\multicolumn{1}{c}{$N_{obs}$}
  &\multicolumn{1}{c}{$N_{ran}$}
  &\multicolumn{1}{c}{Prob.}
  &\multicolumn{1}{c}{$N_{obs}$}
  &\multicolumn{1}{c}{$N_{ran}$}
  &\multicolumn{1}{c}{Prob.} \\
  \hline
$>5$ &5 &  0 & 0.1 &   -- &  0 & 0.1 &   -- &  0 & 0.1 &   -- \\
$>4$ &7 &  0 & 0.1 &   -- &  0 & 0.2 &   -- &  0 & 0.2 &   -- \\
$>3$ &208 &  3 & 3.3 & 64\% &  5 & 4.8 & 52\% &  8 & 6.8 & 36\% \\
\hline
\hline
  \hline
  \end{tabular}
  \label{tab:searchbin}
\end{table*}

\begin{table*}
  \centering
  \caption{Disrupted Binary Candidates}
  \begin{tabular}{rcrrrrrr}
  \hline
  \multicolumn{1}{c}{Gaia EDR3 ID}   &
  \multicolumn{1}{c}{$M_{lim}$} &
  \multicolumn{1}{c}{False}     &
  \multicolumn{1}{c}{$\chi_\varpi^2$} &
  \multicolumn{1}{c}{$\chi_\mu^2$} &
  \multicolumn{1}{c}{$t_{min}$} &
  \multicolumn{1}{c}{$R/R_{SNR}$} & 
  \multicolumn{1}{c}{$\varpi$}  \\
  \multicolumn{1}{c}{} &
  \multicolumn{1}{c}{($M_\odot$)} &
  \multicolumn{1}{c}{(\%)} &
  \multicolumn{1}{c}{} &
  \multicolumn{1}{c}{} &
  \multicolumn{1}{c}{($10^3$~yr)} &
  \multicolumn{1}{c}{} &
  \multicolumn{1}{c}{(mas)} \\
  \hline
5339171053702282112 &3 & 98 &$ 0.591$ &$ 2.083$ &$-4.3 \pm  1.0 $ &$0.27$ &$  0.153 \pm  0.018 $ \\ 
5339171019342516608 &3 & 86 &$ 0.213$ &$ 0.211$ &$-4.5 \pm  0.9 $ &$0.28$ &$  0.156 \pm  0.018 $ \\ 
5339171019342516352 &3 & 99 &$ 3.261$ &$ 0.023$ &$-3.8 \pm  0.8 $ &$0.21$ &$  0.151 \pm  0.019 $ \\ 
5339170847543846656 &3 & 99 &$ 1.164$ &$ 3.933$ &$-3.0 \pm  0.8 $ &$0.14$ &$  0.154 \pm  0.019 $ \\ 
5339170847543834240 &3 &100 &$ 1.058$ &$ 2.133$ &$-1.5 \pm  0.3 $ &$0.02$ &$  0.156 \pm  0.020 $ \\ 
5339170847543832320 &3 & 99 &$ 3.029$ &$ 1.852$ &$-0.9 \pm  0.2 $ &$0.07$ &$  0.180 \pm  0.017 $ \\ 
5339170847543831168 &3 & 99 &$ 0.092$ &$ 0.772$ &$-2.0 \pm  0.4 $ &$0.04$ &$  0.161 \pm  0.019 $ \\ 
5339170808840074752 &3 & 99 &$ 4.507$ &$ 1.471$ &$-1.1 \pm  0.2 $ &$0.06$ &$  0.185 \pm  0.016 $ \\ 
\hline 
\hline 
  \hline
  \end{tabular}
  \label{tab:candbin}
\end{table*}

\begin{table*}
  \centering
  \caption{Properties of Candidates}
  \begin{tabular}{lrccccl}
  \hline
  \multicolumn{1}{c}{Gaia EDR3 ID } &
  \multicolumn{1}{c}{$\chi^2/N_{dof}$} &
  \multicolumn{1}{c}{$\log(T_*/\hbox{K})$} &
  \multicolumn{1}{c}{$\log(L_*/L_\odot)$} &
  \multicolumn{1}{c}{$M_*/M_\odot$} &
  \multicolumn{1}{c}{$\log t$} &
   \\
  \hline
5339171053702282112 &$ 1.00 $ &$  3.672 \pm  0.020 $ &$  1.512 \pm  0.066 $ &$  1.1 $-$  2.3 $ &$ 8.92 $-$ 9.99 $  \\ 
5339171019342516608 &$ 1.50 $ &$  3.648 \pm  0.021 $ &$  1.449 \pm  0.068 $ &$  1.1 $-$  1.8 $ &$ 9.23 $-$ 9.99 $  \\ 
5339171019342516352 &$ 0.48 $ &$  3.646 \pm  0.024 $ &$  0.938 \pm  0.077 $ &$  1.1 $-$  1.3 $ &$ 9.65 $-$ 9.99 $  \\ 
5339170847543846656 &$ 0.81 $ &$  3.646 \pm  0.017 $ &$  0.953 \pm  0.052 $ &$  1.1 $-$  1.2 $ &$ 9.80 $-$ 9.99 $  \\ 
5339170847543834240 &$ 1.16 $ &$  3.823 \pm  0.019 $ &$  1.305 \pm  0.050 $ &$  1.6 $-$  2.3 $ &$< 9.36$           \\ 
5339170847543832320 &$ 1.40 $ &$  3.703 \pm  0.027 $ &$  1.640 \pm  0.081 $ &$  1.0 $-$  2.6 $ &$ 8.75 $-$ 9.99 $  \\ 
5339170847543831168 &$ 1.05 $ &$  3.720 \pm  0.021 $ &$  0.923 \pm  0.057 $ &$  1.1 $-$  1.7 $ &$ 9.33 $-$ 9.99 $  \\ 
5339170808840074752 &$ 1.04 $ &$  3.729 \pm  0.032 $ &$  0.949 \pm  0.101 $ &$  1.1 $-$  2.1 $ &$< 9.99$           \\ 
  \hline
  \end{tabular}
  \label{tab:stars}
\end{table*}

\section{Data and Methods}

\cite{Hughes2001} identified PSR~J1124$-$5916 as a likely pulsar in Chandra X-ray
observations of the SNR G292.0$+$1.8.  This was confirmed by \cite{Camilo2002},
who found it was a 135~msec radio pulsar with a spin down age of only 2900~years.  
Its J2000 position is 11:24:39.0(1) $-$59:16:19(1) based on a Fermi LAT timing
solution (\citealt{Ray2011}). This position is 1\farcs8 from the original Chandra
position.  As noted earlier, \cite{Long2022} measured the proper motion of the
pulsar using several epochs of Chandra observations. They report their final
result as a velocity of $612\pm 152$~km/s at a position angle of $126\pm17$
degrees assuming a distance of $6.2$~kpc.  This corresponds to a proper motion
of $\mu_\alpha = 16.9 \pm 5.6$~mas/year and $\mu_\delta=-12.3\pm 5.9$~mas/year in
the Gaia reference frame. 

G292.0$+$1.8 extends 9.6 arcmin North-South and 8.4 arcmin East-West (\citealt{Park2007})
so we adopt the geometric mean radius of $R_{SNR}=4.5$ arcmin. \cite{Winkler2009} measure
an expansion center of (J2000) 11:24:34.4 $-$59:15:51 and an expansion age of 
$2990 \pm 60$~years based on the proper motions of [OIII] 5007\AA\ emission line
filaments.  The 46~arcsec offset of the pulsar from the expansion center already
implied the pulsar had a high proper motion before its direct measurement.  

\cite{Gaensler2003} estimated a minimum distance of $6.2\pm0.9$~kpc based on the
observed HI absorption profile of the SNR.  \cite{Goss1979} estimated a distance 
of 5.4~kpc based on the optical extinction of the SNR emission, and \cite{Camilo2002}
estimate a distance of 5.7~kpc based on the dispersion measure and the models
of \cite{Cordes2002}. We use the distances of 5.4~kpc and 7.1~kpc to define a
``pseudo-parallax'' for the SNR of $\varpi = 0.16 \pm 0.02$~mas as the mean
of these distances as parallaxes and their separation from the mean as the
uncertainty, rounded to two digits. \cite{Goss1979} estimated that the 
extinction was $E(B-V) \simeq 0.9$~mag.  \cite{Gaensler2003} measure a
hydrogen column density of $N(H)\simeq 3.3 \times 10^{21}$~cm$^{-2}$ which
corresponds to $E(B-V) \simeq 0.6$ for the dust-to-gas ratio of \cite{Bohlin1978}.

We select nearby stars from Gaia EDR3 (\citealt{Gaia2016}, \citealt{Gaia2021}), 
requiring them to have proper motions, all three Gaia magnitudes 
($G$, $B_P$ and $R_P$), $G<19$~mag, $\varpi < 1$~mas and to lie within
within $0.1^\circ$ of the center of the SNR.   This should include all stars 
more massive than approximately $2M_\odot$ for a distance of $6.2$~kpc and
$E(B-V)=0.9$~mag.  We include no restrictions
on the RUWE statistic for the quality of the parallax.  
For the present effort we can ignore the small systematic
uncertainties in the parallax zero point.
We searched the Hipparcos (\citealt{Perryman1997}) and Bright Star 
(\citealt{Hoffleit1995}) catalogs for any bright ($G\ltorder 3$~mag)
stars that would be missing from Gaia and found none.  
At this point we have 3526 stars.

For extinction estimates we used the three dimensional {\tt combined19 mwdust}
models (\citealt{Bovy2016}), which combine the 
\cite{Drimmel2003}, \cite{Marshall2006}
and \cite{Green2019b} models to provide extinction estimates for any sky position
as a function of distance. We
extracted the V band extinction, and use the PARSEC estimates of 
$A_\lambda/A_V$ for the Gaia EDR3 bands and an $R_V=3.1$ extinction law 
to convert the V band extinction to those for the $G$, $B_P$ and $R_P$ bands.
At a distance of 6.2~kpc, the {\tt mwdust} extinction estimate
is $E(B-V) \simeq 0.65$~mag,
and there is a rapid rise in the extinction between 3 and 5~kpc.

We used Solar metallicity PARSEC (\citealt{Bressan2012}, \citealt{Marigo2013},
\citealt{Pastorelli2020}) isochrones to characterize the stars.  
Table~\ref{tab:maglim} gives rough cuts on the extinction-corrected
absolute magnitude $M_G$ as a function of color for selecting stars
more massive than $3M_\odot$, $4M_\odot$ or $5M_\odot$. \cite{Kochanek2021}
also includes the limits for $1M_\odot$ and $2M_\odot$, but for 
G292.0$+$1.8 we are dominated by false positives even for
the rarer higher mass stars.  Because of the distance, most of the
stars lack good parallaxes, so we use the distance corresponding to
the joint parallax estimate between the star and the SNR (see below).
Essentially, this will set the mass scale to the mass a star would have
at the distance to the SNR.  

Our approach is the same as in \cite{Kochanek2021}.
We first simply identify stars whose parallaxes
are consistent with the pseudo-parallax assigned to the SNR by fitting
for a joint parallax and computing the $\chi^2_\varpi$ goodness
of fit.  We initially keep all stars which 
satisfy $\chi^2_\varpi < 9$.  Next we identify all stars
which could intersect the path of the neutron star in the
time interval $-t_{max} < t_m < 0$ with a stellar velocity 
$v_* < v_{max}$ with $t_{max}=10^4$~years and $v_{max}=1000$~km/s. 
We allowed this very high maximum velocity because of the 
discussions of stripping the progenitor in a common envelope
phase (e.g., \citealt{Bhalerao2019}, \citealt{Temim2022}),
which can lead to a compact binary with high orbital speeds.
In practice, the maximum proper motion of the $\chi^2_\varpi<9$
stars is $22.8$~mas/yr in the local standard of rest, or 670~km/s at 6.2~kpc.
This star has a very uncertain parallax ($\varpi = 0.18 \pm 0.22$~mas) and
is probably a slower moving star at a smaller distance. 
This step leaves 1882 stars and
the $N_*$ column of Table~\ref{tab:searchbin} gives the number of 
stars above each $M_{lim}$.

For these stars, we do a joint fit to the proper
motions of each star and the neutron star, introducing the ``true'' proper motions as variables
to be fit to the observed proper motions, and penalize the
separation of the star and the pulsar given the true proper
motions. Thus, the fit statistic in RA for stars $1$ and $2$ is
\begin{eqnarray}
\chi^2_{\alpha,12} &= 
 { \left( \mu_{\alpha,1}-\mu_{\alpha,1}^T\right)^2 \over \sigma_{\alpha,1}^2 }
+{ \left( \mu_{\alpha,2}-\mu_{\alpha,2}^T\right)^2 \over \sigma_{\alpha,2}^2 } \\
&+{ \left( \Delta\alpha_{12} - t(\mu_{\alpha,1}^T-\mu_{\alpha,2}^T)\right)^2 \over
    \sigma_a^2 } \nonumber
\end{eqnarray}
where $\mu_{\alpha,i}$ is the measured proper motion,
$\mu_{\alpha,i}^T$ is the unknown true proper motion,
$t$ is the time, and
\begin{equation} 
   \sigma_a = 10\farcs0 \left( { a_{max} \over 10^4~\hbox{AU} } \right)
        \left( { \varpi \over \hbox{mas}} \right).
    \label{eqn:matchdist}
\end{equation}
sets the scale for the
maximum semi-major axis $a_{max}$ of the binary search.
Note that there is little gain from shrinking $a_{max}$ since it is 
unimportant once the positional uncertainties created by the proper 
motions are larger than $\sigma_a$.  There
is also an analogous contribution $\chi^2_{\delta,12}$ for the
motion in Declination and the combined statistic 
$\chi^2_\mu = \chi^2_{\alpha,12} +\chi^2_{\delta,12}$
can then be minimized.

We optimize $\chi^2_\mu$ with respect to $t^{-1}$.  If
we define $\Delta\mu_{\alpha}=\mu_{\alpha,1}-\mu_{\alpha,2}$,
$\Delta\mu_{\delta}=\mu_{\delta,1}-\mu_{\delta,2}$,
$\sigma_{\alpha,12}^2 = \sigma_{\alpha,1}^2+\sigma_{\alpha,2}^2+\sigma_a^2/t^2$
and
$\sigma_{\delta,12}^2 = \sigma_{\delta,1}^2+\sigma_{\delta,2}^2+\sigma_a^2/t^2$
we can minimize $\chi^2_\mu$ analytically while holding $t$ constant
in $\sigma_{\alpha,12}$ and $\sigma_{\delta,12}$ and then iteratively
update the time used in these uncertainties to deal with the intrinsic
non-linearity.  This then leads to a time of closest approach
\begin{equation}
    t_m = - { \Delta\alpha_{12}^2\sigma_{\delta,12}^2 + \Delta\delta_{12}^2\sigma_{\alpha,12}^2 \over
       \Delta\alpha_{12}\Delta\mu_{\alpha,12}\sigma_{\delta,12}^2+
       \Delta\delta_{12}\Delta\mu_{\delta,12}\sigma_{\alpha,12}^2}
     \label{eqn:time}
\end{equation}
with a propagation of errors uncertainty of
\begin{equation}
\sigma_t = t_m^2 \left( { \sigma_{\alpha,12}^2 \sigma_{\delta,12}^2 \over
         \Delta\delta_{12}^2 \sigma_{\alpha,12}^2 + \Delta\alpha_{12}^2 \sigma_{\delta,12}^2 }\right)^{1/2}
\end{equation}
and a goodness of fit of
\begin{equation}
   \chi^2_\mu = 
   { \left( \Delta\delta_{12} \Delta\mu_{\alpha,12} - \Delta\alpha_{12} \Delta\mu_{\delta,12} \right)^2
     \over \Delta\alpha_{12}^2 \sigma_{\delta,12}^2  + \Delta\delta_{12}^2 \sigma_{\alpha,12}^2 }.
    \label{eqn:binary}
\end{equation}
The same procedure can be used to search for unbound triples and the complete 
expressions are given in \cite{Kochanek2021}.

It is also important to evaluate the probability of having
false positives.  First, we characterize the SNR by scrambling 
the stellar positions relative to their magnitudes, proper motions 
and parallaxes.  The proper motion and parallax errors are largely
determined by the magnitude, so these three quantities cannot be
trivially scrambled.  We randomly assign each star the position of
a different star  and count the fraction of trials which have
closest approaches within 20\% of the SNR radius and joint
fit statistics $\chi^2_\varpi+\chi^2_\mu$
less than several limits.   This test characterizes the general
false positive properties of the SNR given the available data.  

Second, we evaluate the likelihood that each candidate is a false
positive by taking all stars with $\chi^2_\varpi < 9$ and 
within 1~mag of each candidate, putting them at the position of the
candidate and determining the fraction of these trials which 
have a goodness of fit $\chi^2_\varpi + \chi^2_\mu$
fit less than the larger of the value found for the
candidate or 2.706 (90\% confidence).  This tests whether,
for example, a candidate is simply located at a position
where many randomly selected proper motions would still allow
a low fit statistic. This test characterize the false positive
properties of each candidate.  These individual candidate
probabilities can be very different than the more general false positive 
test above if a candidate has a very
different proper motion than the typical star.  

\begin{figure}
\centering
\includegraphics[width=0.50\textwidth]{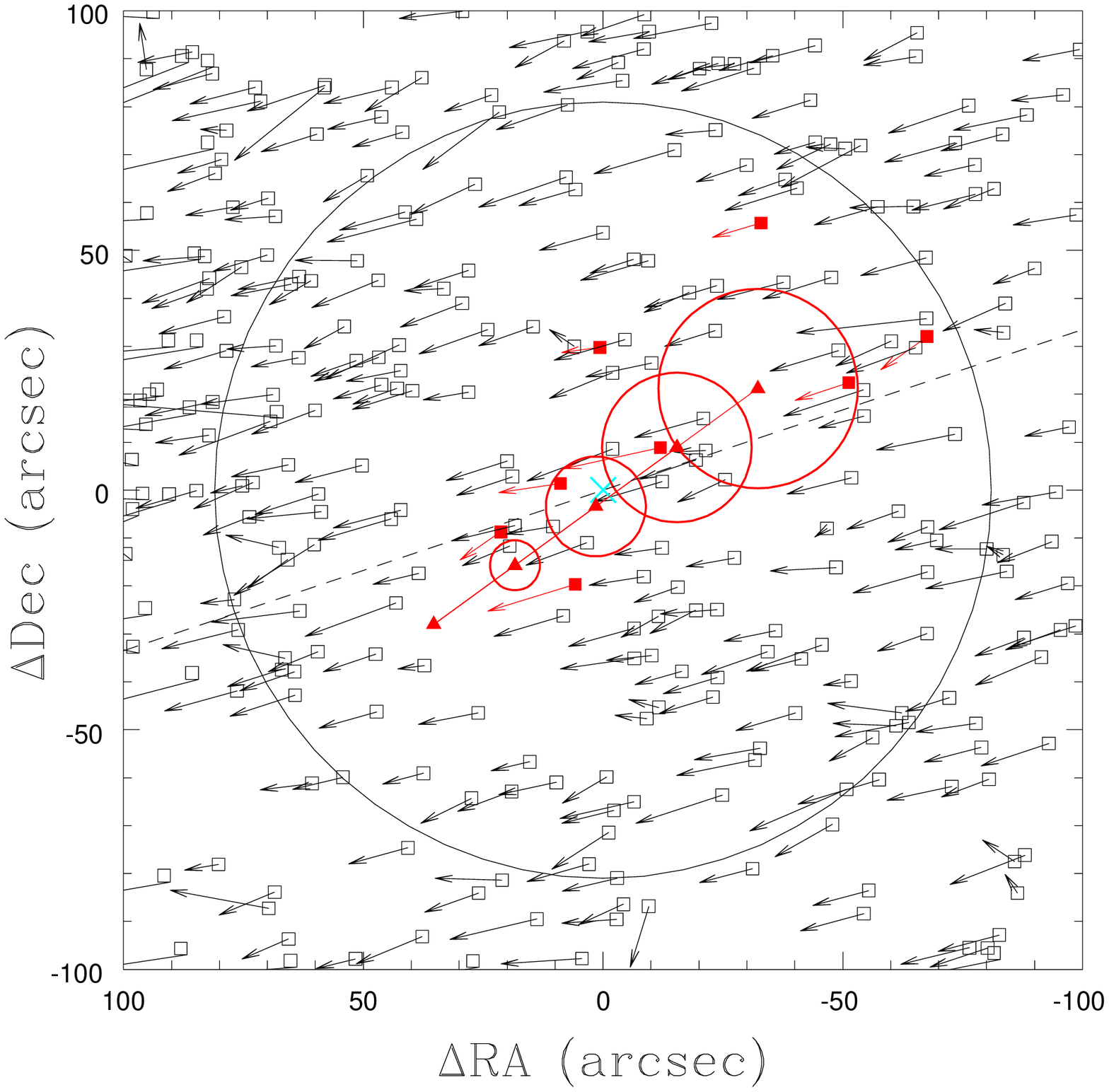}
\includegraphics[width=0.50\textwidth]{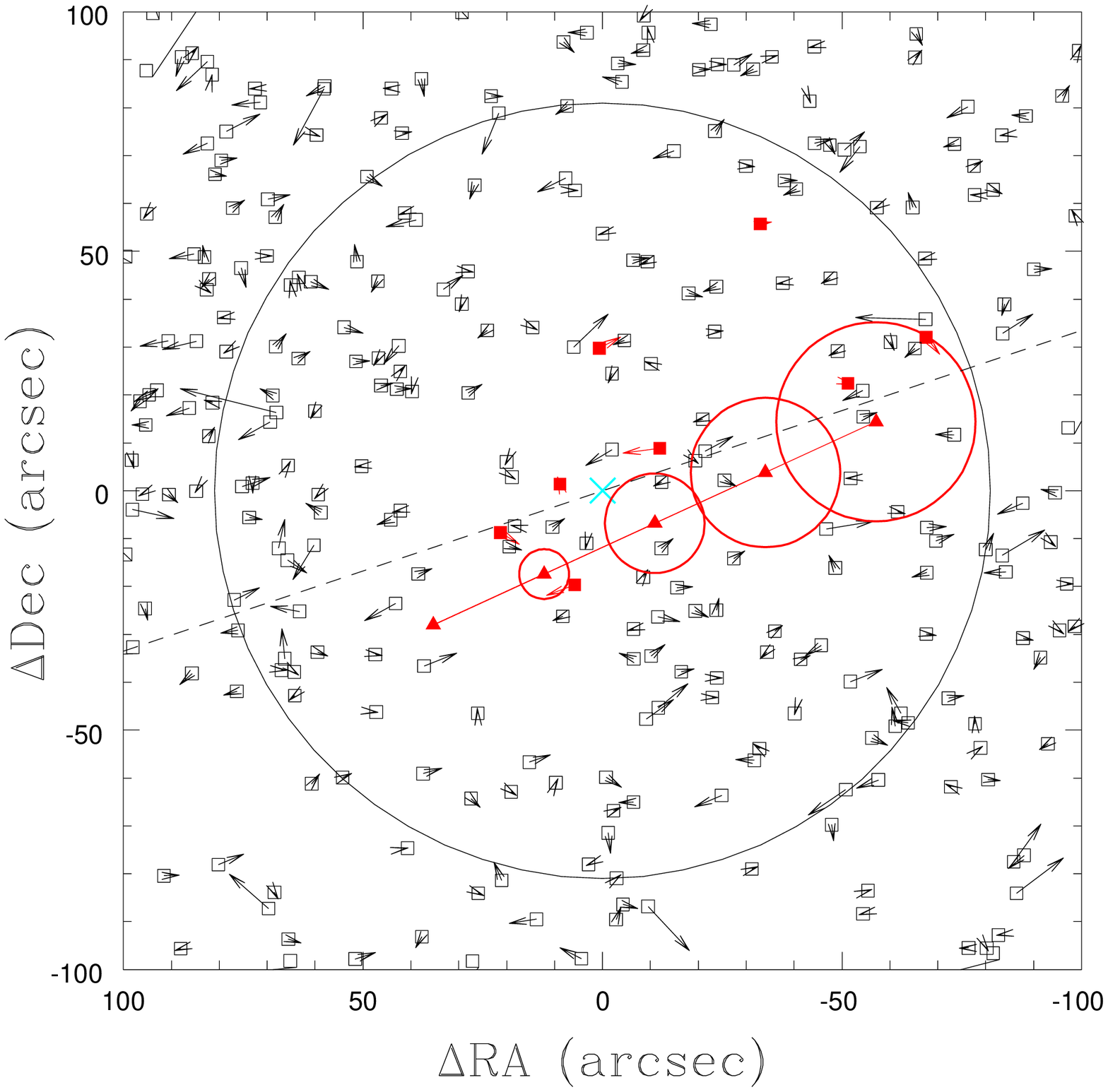}
\caption{
  The motions of the pulsar and its surrounding stars in the Gaia frame (top)
  and the LSR (bottom) defined by the median proper motion of the stars
  with $\chi^2_\varpi <9$.  Squares are the current positions of the stars,
  and the arrows extend to their positions 2000 years ago.  The eight
  binary candidates from Table~\ref{tab:candbin} are in red. The red triangles with the
  growing circles show the position of the pulsar $0$, $1000$, $2000$, $3000$ and $4000$
  years ago with their associated ($1\sigma$) uncertainties shown by the red circles.
  Positions are relative to the \protect\cite{Winkler2009} expansion center, which is
  marked by the cyan cross.  The large black circle corresponds to 30\% of the mean
  radius of the SNR around the expansion center.  We only include stars
  presently in the region or there 2000 years ago.  The dashed line is parallel
  to the Galactic plane, and we see that the Gaia frame proper motions are 
  roughly parallel to the plane because they are dominated by the effects of the
  Galactic rotation curve.
  }
\label{fig:stars}
\end{figure}

\section{Results}
\label{sec:binaries}

Given its space velocity and the non-mention of binarity
in the pulsar timing studies, it is unlikely that the pulsar
can be in a surviving binary. The optical and near-IR counterpart 
to the pulsar is also very faint, with
combined emission from the pulsar and its wind nebula of
$V=24.3$, $R=24.1$, $I=23.1$, $H=21.3$ and $K_s=20.5$~mag
(\citealt{Zharikov2013}). For a distance of $6.2$~kpc and the
{\tt mwdust} extinction estimate of $E(B-V)=0.65$~mag,
these correspond to absolute magnitudes of
$M_V=8.3$, $M_R=8.5$, $M_I=7.9$, $M_H=6.9$ and $M_{Ks}=6.3$~mag,
which limits the mass of any star associated with the pulsar to
$M_*\ltorder 0.2M_\odot$ for a 10~million year Solar metallicity
PARSEC isochrone.  It would be essentially impossible to keep
such a low mass star in a binary after the explosion, so it 
seems safe to conclude that the pulsar is no longer in a binary.  

Fig.~\ref{fig:stars} shows the $\chi^2_\varpi < 9$ stars
superposed on the central region of the SNR both at their present
locations and where they would have been 2000 years ago given
their Gaia EDR3 proper motions.  The coherent motions in the
upper panel are simply
due to the Galactic rotation curve and the Galactic longitude
($\ell = 292^\circ$) of the pulsar.  Using the rough distances and
simply assuming a flat $240$~km/s rotation curve, the predicted
tangential motion is roughly 170~km/s, which is $6$~mas/year.
The stars have mean proper motions of 
($-6.4\pm 1.9$,$1.7\pm 1.3$)~mas, where the uncertainty is
the dispersion not the formal error, and the result changes
little if we tighten the parallax limit.  The median 
proper motions are ($-6.2$, $1.7$)~mas/year. This mean motion of
$6.6$~mas/year is both parallel to the Galactic plane (see
Fig.~\ref{fig:stars}) and a very close match to the expectation from
the simple kinematic model.  There are roughly equal contributions
from our motion and the local standard of rest (LSR) at the pulsar.

Fig.~\ref{fig:stars} also shows the back projection of the 
pulsar to $0$, $1000$, $2000$, $3000$ and $4000$~years ago
in the Gaia kinematic frame of reference
along with the associated uncertainty given the estimated
errors on the pulsar proper motions from \cite{Long2022}.
The
errors on the pulsar proper motions completely dominate
the problem: the Gaia EDR3 proper motion uncertainties 
are roughly 50 times smaller than those of the pulsar, and
a $10^4$~AU orbit only subtends only 1\farcs6 at 6.2~kpc (Eqn.~\ref{eqn:matchdist}).
A star in Fig.~\ref{fig:stars} would be good 
unbound binary candidate with an age of 2000 years if 
the head of its arrow lies in the corresponding pulsar error
circle. 

Fig.~\ref{fig:stars} also shows the motions in the local
standard of rest defined by the median proper motions of
the stars with $\chi^2_\varpi < 9$.  In this frame, the
motions of the stars over 2000~years are generally 
negligible compared to the uncertainties in the back
propagated position of the pulsar.  The pulsar proper
motion is now ($23.1$, $10.6$)~mas/year instead of
($16.9$, $12.3$)~mas/year with the total
increasing to $25.4$~mas/year from $20.9$~mas/year.
This increases the velocity from $612$~km/s to 
$744$~km/s at a distance of $6.2$~kpc. It is now 
quite difficult to reconcile the \cite{Winkler2009}
expansion center of the SNR, the pulsar motions and 
an age close to $3000$~years.  An age of $2000$~years 
remains acceptable.

The Gaia frame version of Fig.~\ref{fig:stars} is not
fully consistent because we have presented it as if the
expansion center is fixed in this frame.  In reality,
if the explosion is spherical and the ISM is roughly
comoving with the stars, then the velocity 
of the expansion center is the 
same as the pre-SN velocity of the
progenitor star. If any pre-SN orbital velocity was small,
then in the Gaia frame of Fig.~\ref{fig:stars}, the 
expansion center is moving to the South-East with roughly
the median proper motions of the stars.
Including this would reconcile 
the apparently different positions of the pulsar relative 
to the expansion center in the Gaia and LSR
frames shown in Fig.~\ref{fig:stars}.  

We assume
the expansion center of the SNR is at rest with respect to the LSR,
and so give it the median proper motion
of the stars when we compute the distance of stars from 
the explosion center as a function of time.  While certainly wrong at some level,
it is a better assumption than keeping it at rest in the
Gaia frame given that there are both large differences between the 
Gaia frame and the LSR and that these differences have nothing
to do with any physics leading to the expansion center having
a motion relative to the LSR.  Compared to \cite{Kochanek2021}
we also used a maximum fractional distance of the closest 
approach point from the expansion center of of $R/R_{SNR}<0.3$ instead of $0.2$. 
 
Table~\ref{tab:searchbin} provides the general false positive character
of G292.0$+$1.8 based on randomly mixing the coordinates and proper
motions.  It counts the number of real and random candidates which
have have $R/R_{SNR}<0.3$ at the time of closest approach and
$\chi^2_\varpi+\chi^2_\mu<2.71$, $3.84$, or $6.64$, corresponding
to probabilities that $90\%$, $95\%$ and $99\%$  of systems should
have smaller fit statistics.  Clearly, the probability of candidates
being false positives is high given the proper motion statistics and
their uncertainties.  In essence, the proper motions of the stars
are all pretty similar, particularly compared to the uncertainties
in the pulsar proper motions. So using the proper motions
of a randomly selected star has very little effect on whether 
a given stellar coordinate will produce a 
good candidate.  Visually, if a star in Fig.~\ref{fig:stars} can lie
inside the pulsar error circle at some time, it is highly likely to
do so no matter which arrow is assigned to it.   

All the individual candidates with $M_{lim}>3M_\odot$, 
$\chi^2_\varpi+\chi^2_\mu<6.64$ and $R/R_{SNR}<0.3$ are listed in 
Table~\ref{tab:candbin}, along with their individual false positive rate 
estimates, the times of closest approach, the distances from the expansion
center $R/R_{SNR}$ at that time and the joint parallaxes of the star
and the SNR.  These eight candidates are
marked in Fig.~\ref{fig:stars}. As discussed in \S2, the individual false positive rate
is the probability that the star would still be a candidate if we 
give it the proper motion of a randomly selected star within 1~mag
of its observed magnitude.  These are all essentially 100\%, which is
again a manifestation of the similarity of the stellar proper motions relative
to the uncertainties in the pulsar proper motion.  There are increasing
numbers of candidates for larger ages because the area of the pulsar 
error circle grows quadratically with time (Fig.~\ref{fig:stars}), limited only
by when significant fractions of the error circle begin to be too
far from the expansion center ($R/R_{SNR}>0.3$)
at the time of closest approach.  If we keep the expansion center
position fixed in the Gaia frame, the number of candidates increases 
significantly because it takes longer for the pulsar to approach the 
$R/R_{SNR}<0.3$ limit.  We used fairly generous age, goodness of fit
and $R/R_{SNR}$ limits for Table~\ref{tab:candbin}, and we could
reasonably prune the list by tightening the limits.  However, 
even the remaining candidates would still be ambiguous because of
the high false positive probabilities.  

We can conclude, however, that any former stellar companion cannot have
been a massive star.   To better quantify this than the crude $M_{lim}$
values, we made spectral energy distribution (SED) models using \cite{Castelli2003}
stellar atmospheres for the eight stars in Table~\ref{tab:candbin}
using AllWISE W$_1$/W$_2$ (\citealt{Cutri2014}), 2MASS JHK$_s$ (\citealt{Skrutskie2006}), 
and ATLAS Refcat griz (\citealt{Tonry2018}) photometry.  We use either the
reported photometric error or a minimum error of 0.1~mag to account for 
systematic uncertainties (e.g., metallicity etc.).  We fixed the distance
to $6.2$~kpc and used an extinction prior of $E(B-V)= 0.65 \pm 0.10$ based
on the {\it mwdust} extinction estimate for that distance.     

Table~\ref{tab:stars} gives the results.  They are all relatively cool
stars (4700~K to 6400~K) with luminosities of $10L_\odot$ to 
$62 L_\odot$.  The fitted extinctions are broadly consistent with
the prior, although most stars want modestly lower extinctions.
Forcing the extinction to be $E(B-V)=0.65$~mag does not significantly
alter the results.  We matched the luminosities and temperatures
to Solar metallicity PARSEC isochrones using minimum luminosity
and temperature uncertainties of $0.1$ and $0.05$~dex, respectively, finding
the mass and age ranges consistent ($\chi^2<4$) with the luminosity
and temperature estimates.  The stars appear to be moderately evolved $1.0$ to 
$2.6 M_\odot$ stars, although we suspect that they are mostly lower luminosity
main sequence stars with overestimated luminosities from placing them
at $6.2$~kpc.  That the masses are lower than the $M_{lim}>3M_\odot$
limit is not an issue -- these limits are designed to be very conservative.

\begin{table*}
  \centering
  \caption{Constraints on Binaries}
  \begin{tabular}{lcccc}
  \hline
  \multicolumn{1}{c}{Case} &
  \multicolumn{2}{c}{Non-Interacting Incomplete}  &
  \multicolumn{2}{c}{Non-Interacting Complete  }  \\
  &
  \multicolumn{1}{c}{Median} &
  \multicolumn{1}{c}{90\% Confidence} &
  \multicolumn{1}{c}{Median} &
  \multicolumn{1}{c}{90\% Confidence} \\
  \hline
  Not Binary at Death     &73.5\% &54.6\%87.2\% &77.4\% &58.4\%-89.7\% \\
  Bound Binary            &13.4\% &5.2\%-26.3\%  &9.2\%  &3.5\%-19.2\%  \\
  Interacting Binary      & 9.8\% &3.1\%-21.5\%  &5.6\%  &1.8\%-12.8\%  \\
  Non-Interacting Binary  &       &$<$8.2\%      &       &$<$8.6\%      \\
  Unbound Binary          &11.6\% &2.6\%-29.2\%  &12.2\% &2.7\%-30.6\%  \\
  \hline
  \end{tabular}
  \label{tab:binary}
\end{table*}

\section{Discussion}
\label{sec:conclude}

The current proper motion uncertainties for the pulsar prevent us from
providing an unambiguous answer for the pre-supernova binary status of
PSR~J1124$-$5916.  There are 8 candidate stars with various levels of
plausibility, all of which have very high false positive properties.  
We suspect, but cannot prove, that they are all false positives. 
The \cite{Long2022} proper motions measurements used Chandra data spanning 
2006 to 2016, so it is already feasible to reduce the proper motion
uncertainties by $50\%$, which would significantly reduce the false positive
problem.  There are no candidates for a fully disrupted triple 
system. 

All eight of the stars are relatively low mass, $<3M_\odot$, so
none could be a significant (relative to the initial mass of a
neutron star progenitor) mass gainer.  They also all have 
relatively low proper motions (see Fig.~\ref{fig:stars}), so none are candidates for
a disrupted, short period post-common envelope binary.  If
the progenitor of PSR~J1124$-$5916 was a highly stripped
star, as argued by \cite{Bhalerao2019} and \cite{Temim2022},
it seems unlikely that it was stripped by binary interactions.  

We can also update the binary status probabilities of neutron star producing
SN from \cite{Kochanek2021}.  With the addition of this system, we have 10
systems which were not binaries with $M \gtorder 3M_\odot$ companions at death, 1 unbound 
binary, 2 interacting binaries (this assume SS~433 is a black hole system), 
and 11 which are not bound binaries but
could be unbound binaries.  If $f_n=1-f_u-f_b$ is the fraction that are
not binaries at the time of the SN, $f_u$ is the fraction producing unbound
binaries, fraction $f_b=f_i+f_p$ are bound binaries after the SN with 
$f_i$ interacting and $f_p$ non-interacting (passive), then the 
joint multinomial probability is
$$ f_n^{10} \left( 1 - f_b \right)^{11} f_u f_i^2 
  \label{eqn:prob}
$$
from which we can compute the median and 90\% confidence limits on the
terms given in Table~\ref{tab:binary}.  There is an ambiguity from
\cite{Kochanek2019} about whether interacting binaries are so easy
to detect that the two in the sample were all the interacting binaries
in the parent sample of 49 SNRs they considered.  If this is true,
then the joint multinomial probability should be 
$$ f_n^{10} \left( 1 - f_b \right)^{11} (1-f_i)^{20} f_u f_i^2. 
$$
This case is given in the ``non-interacting complete'' columns.  Adding
a single system does not change the the from \cite{Kochanek2021}
results significantly, but it shows that steady progress can be made 
in observationally determining all of these parameters.  Adding this
system, there are 11 systems which are not fully unbound triples, so
of systems not leaving a bound binary, $<17.5\%$ can be fully unbound
triples at 90\% confidence.

As noted earlier, there are roughly $300$ known Galactic SNRs, and this
statistical analysis makes use of only $\sim 10\%$ of them.  If the 
statistics used in Eqn.~\ref{eqn:prob} could simply be doubled, the
constraints on binary properties improve dramatically.  For example,
if we literally simply double each exponent in Eqn.~\ref{eqn:prob},
the uncertainty in the fraction of systems which are binaries at
death drops from 55\%-87\% (Table~\ref{tab:binary}) to 65\%-88\%.
and the error on the fraction in bound binaries drops from
5\%-26\% to 5\%-19\%.  Much of this can be done with added studies
like \cite{Long2022} to measure the proper motions of known compact
objects. Proper motions can also help to identify the compact
objects in SNRs where there are multiple X-ray sources since active
galactic nuclei, the primary contaminant, have no proper motions.
It is, however, necessary to obtain such observations with Chandra
because of its superior angular resolution to other existing and
planned X-ray missions.  As we have already noted in 
\cite{Kochanek2021} and \cite{Kochanek2022} combining Gaia stellar 
distances with searches for high velocity stellar absorption features
from the SNR, as done by \cite{Cha1999} for the Vela remnant, should
provide a means of obtaining more accurate distances to SNRs.  Poor
SNR distances are one of the main contributors to false positives.   

\section*{Acknowledgments}

CSK is supported by NSF grants AST-1814440 and AST-1907570.  
This work has made use of data from the European Space Agency (ESA) mission
{\it Gaia} (https://www.cosmos.esa.int/gaia), processed by the {\it Gaia}
Data Processing and Analysis Consortium (DPAC,
https://www.cosmos.esa.int/web/gaia/dpac/consortium). Funding for the DPAC
has been provided by national institutions, in particular the institutions
participating in the {\it Gaia} Multilateral Agreement.
 This research has made use of the SIMBAD database,
operated at CDS, Strasbourg, France

\section*{Data Availability Statement}

All data used in this paper are publicly available.

\end{document}